\documentclass[12pt,a4paper]{conference}

\usepackage{fancyhdr}
\usepackage{graphicx,amsmath,amssymb,cite}
\usepackage{multind}
\makeindex{author} \makeindex{subject}

\newcommand{\beq}{\begin{equation}}
\newcommand{\eeq}[1]{\label{#1}\end{equation}}
\newcommand{\eeqn}{\end{equation}}

\newcommand{\beqa}{\begin{eqnarray}}
\newcommand{\eeqa}[1]{\label{#1}\end{eqnarray}}
\newcommand{\eeqan}{\end{eqnarray}}

\let\bar=\overbar

\newcommand{\Dslash}{\not{\hbox{\kern-4pt $D$}}}
\newcommand{\dslash}{\not{\hbox{\kern-2pt $\del$}}}

\newcommand{\msb}{{\bar{\ssstyle M \kern -1pt S}}}

\begin{document}
%%%%%%%%%%%%%%%%%%%%%%%%%%%%%%%%%%%%%%%%%%%%%%%%%%%%%%%%%%%%%%%%%%%%%%%

\Chapter{Lattice Approach to Light Scalars}
           {Lattice Approach to Light Scalars}{Craig McNeile}
\vspace{-5 cm}\includegraphics[width=6 cm]{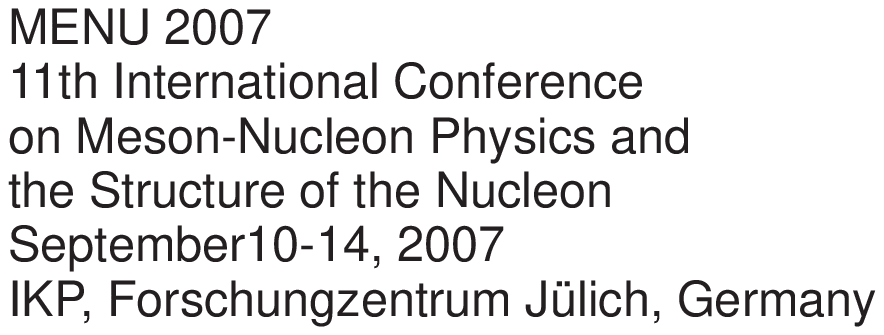}
%%\vspace{-6 cm}\includegraphics[width=6 cm]{Header.eps}
%\bigskip\bigskip
\vspace{4 cm}

\addcontentsline{toc}{chapter}{{\it N. Author}} \label{authorStart}
%%%%%%%%%%%%%%%%%%%%%%%%%%%% NEW SWITCHES %%%%%%%%%%%%%%%%%%%%%%%%%%%%%%

\begin{raggedright}

{\it Craig McNeile\footnote{Current address.}} \\
Department of Physics and Astronomy\\
The Kelvin Building\\
University of Glasgow\\
Glasgow G12 8QQ\\
U.K.\\
\bigskip\bigskip

%%%%%%%%%%%%%%%%%%%%%%%%%%%%%%%%%%%%%%
%%%%%%%%%%%
%%%%%%%%%%%  Repeat for second author
%%%%%%%%%%%
%%%%%%%%%%%%%%%%%%%%%%%%%%%%%%%%%%%%%%
\end{raggedright}

\begin{center}
\textbf{Abstract}
\end{center}
I report on lattice QCD calculations that study
the properties of the $a_0$ and $f_0$
mesons.

\section{Introduction}  \label{se:section}

I review the recent lattice results
for the light $J^{PC}$ = $0^{++}$ scalar mesons.
The interpretation of many $0^{++}$ mesons 
in terms of quark and glue degrees of freedom 
is still not 
clear~\cite{Liu:2007hm,Pennington:2005am}. 
The $0^{++}$ mesons potentially contain glueball, tetraquark,
meson molecule or even quark-antiquark degrees of freedom.
I have recently written a review~\cite{McNeile:2007fu}
of light meson spectroscopy from lattice QCD,
that contains more detail on many of the
topics covered here.

\subsection{Background to lattice QCD}

The physical picture behind lattice QCD calculations
is that an
interpolating operator creates a hadron in the QCD vacuum and after a
specific time interval the hadron is destroyed. The choice of
interpolating operator is particularly important for hadrons where it
is not clear how the hadron is built out of quarks and
gluons.

For example, to create a light flavour singlet $0^{++}$ hadron,
possible interpolating operators are
\begin{eqnarray}
O_1 & = &  \overline{q}q  \label{eq:standard} \\
O_2 & = &  \overline{q}\gamma_5 q \overline{q} \gamma_5 q
\label{eq:hybrid} \\
O_3 & = & U_{plaq}
\label{eq:glueball}
\end{eqnarray}
where $U_{plaq}$ is a spatial 
plaquette of gauge fields
with $0^{++}$ symmetry, and 
$q$ is a light quark operator. 

The majority of recent lattice QCD calculations
include the dynamics of sea quarks and have pion masses as low as 300
MeV~\cite{McNeile:2007fu}. 
The results I will present for scalar mesons largely use 
the last generation of
lattice QCD calculations that are quenched or dynamical QCD
calculations with pion masses 
above 500 MeV~\cite{Allton:2001sk,AliKhan:2001tx}.

There are a number of reasons that lattice calculations of the light
scalar mesons are challenging.  The lattice QCD correlators for scalar
mesons are more noisy than for $\rho$ and $\pi$ mesons.  The light
scalar mesons decay via S-wave decays, and current lattice QCD
calculations are in the quark mass regime where some decay channels
to two mesons are open.

Eventually, the issue of dealing with resonances 
in lattice QCD will be dealt with by L\"{u}scher's
formalism~\cite{Luscher:1991cf} that produces scattering
phase shifts. 
This year L\"{u}scher's technique for resonances
was applied to the $\rho$
meson for the first time,
by the CP-PACS collaboration~\cite{Aoki:2007rd}.

\subsection{The flavour non-singlet $0^{++}$ and $0^{+}$ mesons.}

Although I am going to loop through the lattice results for the
lightest $0^{++}$ and $0^{+}$ mesons, it is important to classify
the states into SU3 multiplets or some other classification based on
tetraquarks for example.

In the PDG the lightest strange-light $0^{+}$ meson is 
the $K_0(1430)$~\cite{Yao:2006px}. There have also been
claims that experimental data is consistent with 
$0^{+}$ $I=1/2$ meson called the $\kappa$ with
a mass of 660 MeV~\cite{DescotesGenon:2006uk}.
The existence of the $\kappa$ is controversial,
see~\cite{Wada:2007cp,Liu:2007hm,Pennington:2005am}
for a discussion. 

%%
%% strange-light
%%

In table~\ref{tb:zeroppsummary}
I collect results for the mass of the lightest 
$0^+$ $\overline{s}q$ meson from 
lattice QCD calculations.
%%%
\begin{table}[tb]
\centering
\begin{tabular}{|c|c|c|} \hline
Group   &  $n_f$  &  $m_{K_0}$ GeV \\ \hline
Prelovsek et al.~\cite{Prelovsek:2004jp} & 2 & $1.6 \pm 0.2$ \\ 
McNeile and Michael~\cite{McNeile:2006nv} & 2 &  $ 1.1 - 1.2$   \\ 
Mathur et al.~\cite{Mathur:2006bs} & 0 & $1.41 \pm 0.12$ \\
SCALAR~\cite{Wada:2007cp} & 0  &  $\sim 1.7$  \\
\hline
\end{tabular}
\caption{Lightest strange-light $0^{+}$ meson from lattice QCD.}
\label{tb:zeroppsummary}
\end{table}
%%
%%%
%%%  MILC
%%%
The lattice results in table~\ref{tb:zeroppsummary}
are consistent with experimental
mass of the $K_0^\star$(1430), but mostly miss 
the controversial  $\kappa$ particle.
All the lattice calculations used $\overline{s}q$
interpolating operators, so may have missed the 
$\kappa$ state, if it is mostly a 
tetraquark state, with no overlap with
$\overline{s}q$ interpolating operators.
%%%
%%%  a0 masses
%%%

Experimentally the lightest $I=1$ $0^{++}$ mesons are 
the $a_0(980)$ and the $a_0(1450)$~\cite{Yao:2006px}. 
There have been
speculations that the $a_0(980)$ meson is a molecule
or tetraquark state~\cite{Liu:2007hm,Pennington:2005am}, 
so it is interesting to see whether
lattice QCD calculations with $\overline{q}q$ interpolating
operators couple to the $a_0(980)$ meson.
In quenched QCD there 
is a ghost contribution~\cite{Bardeen:2001jm}, 
due to the $\eta \pi$ contribution,
to the scalar correlator that 
needs to be subtracted off the lattice data.
I collect together some recent results for the 
mass of the light $0^{++}$ meson from lattice QCD
in table~\ref{tab:a0quenched}. I only include quenched 
data where the $\eta \pi$ contribution has been corrected
for~\cite{Bardeen:2001jm}.

\begin{table}[tb]
\centering
\begin{tabular}{|c|c|c|}
\hline
Group         &   $n_f$  & $m_{a_0}$ GeV \\ \hline
Bardeen at al.~\cite{Bardeen:2001jm} & 0  & $1.34(9)$ \\
Burch et al.~\cite{Burch:2006dg} &  0  & $\sim 1.45$ \\
Hart et al.~\cite{Hart:2002sp}  & 2P & $1.0(2)$ \\
Prelovsek et al.~\cite{Prelovsek:2004jp} & 2  & $1.58(34)$ \\
Prelovsek et al.~\cite{Prelovsek:2004jp} & 2P & $1.51(19)$ \\
Mathur et al.~\cite{Mathur:2006bs} & 0  & $1.42(13)$  \\
\hline
  \end{tabular}
\caption{A collection of results from lattice QCD for the 
mass of the lightest non-singlet $0^{++}$ meson.
The P stands for a partially quenched analysis.
} 
\label{tab:a0quenched}
\end{table}

McNeile and Michael~\cite{McNeile:2006nv}, 
in an unquenched
lattice QCD calculation 
focused on the mass difference (in the hope that
systematics cancel), 
between the $1^{+-}$ and the $0^{++}$ mesons. 
The lattice calculation used gauge configurations from
UKQCD~\cite{Allton:2001sk} and CP-PACS~\cite{AliKhan:2001tx}.
The mass of the light $1^{+-}$ state
was always higher than the $0^{++}$ meson.
The final result was
\mbox{ $m_{b_1} - m_{a_0} = 221(40) $ MeV},
compared to the experimental result of 245 MeV.
Lang et al. recently 
reported masses for the lightest flavour non-singlet
$0^{++}$  consistent with the mass of the $a_0(980)$ meson,
from an unquenched lattice QCD calculation using
chirally improved fermions~\cite{Frigori:2007wa}.

The previous lattice QCD calculations, discussed in this section, were
in a regime where the quark masses were large enough that the decay
$a_0 \rightarrow \eta \pi$ was forbidden. Now I discuss the new
lattice QCD calculations where the decay $a_0 \rightarrow \eta \pi$ is
energetically allowed.

The MILC collaboration~\cite{Bernard:2001av}
originally 
claimed that they had evidence 
for $a_0$ decay to $\pi\eta$ from
their calculations with improved staggered fermions.
Other decays are discussed in~\cite{Aubin:2004wf}.
Later work by the MILC~\cite{Aubin:2004wf} 
and UKQCD~\cite{Gregory:2005yr} collaborations showed that the 
lightest state in the flavour non-singlet $0^{++}$ channel 
was actually below the $\pi\eta$ threshold,
with improved staggered fermions.
This was puzzling, because experimentally 
the $a_0 \rightarrow \pi\pi$ 
decay is forbidden by G parity.

In~\cite{Prelovsek:2005rf}, Prelovsek explained 
the behaviour of the flavour non-singlet $0^{++}$
correlator with improved staggered fermions using
staggered chiral perturbation theory.
Bernard, DeTar, Fu, and Prelovsek~\cite{Bernard:2007qf} 
extended the original analysis by Prelovsek, and
also applied it to the flavour singlet $f_0$ meson.
A larger study, with more sea quark
masses, is required to say something specific
about the mass of the lightest $a_0$ meson.

%%
%% new results from ETMC 
%%

\begin{figure}
%%% PROB1
\centering
\includegraphics[%
  scale=0.35,
%% angle=270,viewport=0 0 600 700,
 angle=270,
  origin=c,clip]{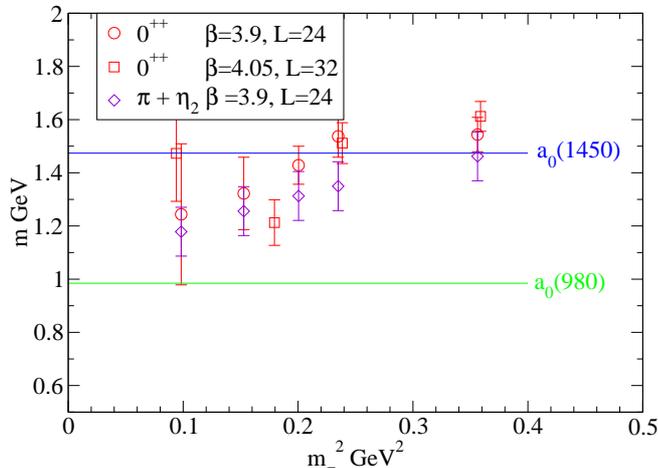}
\vspace{-2.0cm}
\caption{Mass of lightest state in $0^{++}$ channel with $\pi\eta_2$
decay threshold.}
\label{fig:ETMCa0}
\end{figure}

The ETM collaboration have preliminary results
for the mass of the light $0^{++}$ meson from
a $n_f$=2 unquenched lattice QCD calculation with
twisted mass 
fermions~\cite{Boucaud:2007uk,Michael:2007vn,Urbach:2007rt}. 
In figure~\ref{fig:ETMCa0} I plot
the mass of the light $0^{++}$ meson and the 
$\pi + \eta_2$ decay threshold as a function of the square
of the pion mass. There was a bug in the original preliminary analysis
of the $a_0$ masses from ETMC, however the plot 
in figure~\ref{fig:ETMCa0} is from arXiv:0906.4720
and is correct.
The mass of the $\eta_2$ was computed 
by Michael and Urbach~\cite{Michael:2007vn}.
Figure~\ref{fig:ETMCa0} shows 
some evidence for the mass of the $0^{++}$ tracking
the $\pi + \eta_2$ threshold, or at least for it
being an open decay channel.

Some caution is required in the interpratation of the results,
we are only just starting to deal with mesons with
open decay channels in unquenched lattice QCD calculations. 
There is a of order 250 MeV
difference between the mass of the lightest 
flavour singlet pseudoscalar
meson in lattice QCD calculations with $n_f=2$ and 
$n_f=2+1$ sea quark flavours~\cite{McNeile:2007fu}
and this will be important for the decay thresholds.
%%

%% leptonic decay constant
%%
There are other quantities, other than masses, 
that can help determine
the quark and glue content of scalar mesons.
For example, Narison~\cite{Narison:2005wc} 
proposed to use the leptonic decay
constant of the non-singlet $0^{++}$ mesons to determine
the structure of the $a_0$ meson. The $f_{a_0}$
decay constant of the light flavour non-singlet $0^{++}$ meson
has been computed using unquenched 
lattice QCD~\cite{McNeile:2006nv}.
\begin{equation}
\langle 0 \mid \overline{q} q | a_0 \rangle
= M_{a_0} f_{a_0}
\label{eq:decayDEFN}
\end{equation}
See~\cite{McNeile:2006nv,Diehl:2001xe,Cheng:2005nb} 
for a further 
discussion of this decay constant and the connection 
with the electroweak current. 

A molecule of two mesons should have a very small 
"wave-function" at the origin, hence $f_{a_0}$ should be small.
The definition of $f_{a_0}$ is similar to that of 
the pion decay constant. Hence we mean "small" relative to 
130 MeV. The other measured decay constants of 
pseudoscalar mesons are within a factor of 2.5 to 
the pion decay constant~\cite{Yao:2006px}.
The only exception is the decay constant of the 
$\pi(1300)$ that is 
suppressed~\cite{Holl:2004fr,McNeile:2006qy}.
A large value for decay constant $f_{a_0}$ 
does not rule out a $\overline{q}\overline{q}qq$
multi-quark meson.

Using 
gauge configurations from UKQCD and CP-PACS,
McNeile and Michael computed $f_{a_0} \sim 480 $ MeV.
Sum rule and model estimates find  $f_{a_0}$
in the range 290 to 440 
MeV~\cite{Maltman:1999jn,Shakin:2001sz,Narison:2005wc,Cheng:2005nb}.
The $f_{a_0}$ decay constant depends on the scale and this
should be specified for a more detailed comparison.
%%%

Computing the decay width of a hadron is 
also very a valuable way of identifying a 
state on the lattice. In~\cite{McNeile:2006nv},
it was reported that the experimental hadron coupling
for the decays $a_0(980) \rightarrow K \overline{K}$ 
and
$a_0(1450) \rightarrow K \overline{K}$  were
0.9 and 0.5 respectively.  
A lattice calculation~\cite{McNeile:2006nv} 
found that the 
lightest hadron in the $0^{++}$ correlator
had a coupling to $K\overline{K}$ of $\approx 1$,
thus providing additional evidence that 
the lightest state was the $a_0(980)$.

Pennington~\cite{Pennington:2007yt} has recently
extracted the two photon decay width of the $\sigma$
from experiment to be
$\Gamma (\sigma \rightarrow \gamma \gamma) \sim $ 4 keV.
Pennington notes that value of 
$\Gamma (\sigma \rightarrow \gamma \gamma)$ can depend
quite sensitively on the quark content of the 
$\sigma$~\cite{Pennington:2007yt}.
Recently a 
formalism to compute two photon widths on the lattice
has been developed~\cite{Dudek:2006ut}.
Dudek and Edwards have
computed $\Gamma(\chi_0 \rightarrow \gamma \gamma$) = $2.4 \pm 1.0$ keV,
from a quenched QCD calculation~\cite{Dudek:2006ut}. 
It would be interesting to do a 
similar calculation for light scalar mesons.

\subsection{Flavour singlet $0^{++}$ mesons}

The spectrum of the light flavor singlet $0^{++}$ mesons
is where the $0^{++}$ glueball is thought to be hiding out.
The lightest flavor singlet $0^{++}$ mesons
listed in the PDG~\cite{Yao:2006px} are:  
$f_0(600)$, $f_0$(980), $f_0$(1370), $f_0$(1500), and $f_0$(1710).
There are claims that the $f_0$(980) is a molecule 
or tetraquark~\cite{Yao:2006px}, so it may not couple to
$\overline{q}q$ interpolating operators.

Morningstar and Peardon~\cite{Morningstar:1999rf}
obtained $M_{0^{++}}$ = 1730(50)(80) MeV
for the mass of the lightest $0^{++}$ glueball from
quenched QCD.
Chen et al. ~\cite{Chen:2005mg} recently
found $M_{0^{++}}$ = 1710(50)(80) MeV.
The quark model predicts that there should only
be two $0^{++}$ mesons between 1300 and 1800 MeV, so
if the mixing between the glueball and $\overline{q}q$
operators is weak, then the $0^{++}$ glueball is hidden inside
the $f_0(1370)$, $f_0(1500)$ and $f_0(1710)$ mesons.

Weingarten and Lee~\cite{Lee:1999kv}
used quenched lattice QCD to estimate the 
mixing matrix 
between the glue and $\overline{q}q$ states.
Weingarten and Lee~\cite{Lee:1999kv} 
predicted that the $f_0(1710)$ meson was 74(10)\% $0^{++}$
glueball, and hence the mixing between the $0^{++}$ glueball
and $\overline{q}q$ states was weak. 

There are claims~\cite{Mennessier:2007wk}
 that continuum 
phenomenology is more consistent with a
a sizable contributions from 
the $0^{++}$ glueball
to the $f_0(600)$ and $f_0(980)$ mesons.

%%
%% history of unquenched
%%
The SESAM collaboration studied the glueball spectrum
on unquenched lattices~\cite{Bali:2000vr}.
McNeile and Michael studied the light $0^{++}$ spectrum
with unquenched QCD~\cite{McNeile:2000xx} at a coarse
lattice spacing
and found the mass of the lightest flavour singlet
$0^{++}$ meson was very light.
Using $0^{++}$ glueball operators, 
Hart and Teper~\cite{Hart:2001fp} found that 
\begin{equation}
M_{0^{++}UNquenched} = 0.85(3) M_{0^{++}Quenched}
\end{equation}
at a fixed lattice spacing of 0.1 fm.
The UKQCD collaboration~\cite{Gregory:2005yr}
separately studied $0^{++}$ glueball 
and  $0^{++}$ $\overline{q}q$
operators on improved staggered gauge configurations,
however higher statistics and an analysis similar to the one
by Bernard et al. is required~\cite{Bernard:2007qf}.

Unfortunately, the existing unquenched lattice QCD calculations
of the flavour singlet $0^{++}$ mesons don't have the 
range of lattice spacings where a continuum extrapolation
can be attempted.
In quenched QCD it was found that the lattice spacing 
dependence of the mass  of the $0^{++}$ 
glueball was strong. 
The use of a Symanzik improved gauge action by 
Chen et al.~\cite{Chen:2005mg}
and, Morningstar and Peardon~\cite{Morningstar:1999rf},
produced a slightly smaller slope with lattice spacing
of the scalar $0^{++}$ glueball mass, than for calculations
that used the Wilson plaquette action. This 
is relevant to unquenched calculations, because 
any suppression of the mass of the flavour singlet $0^{++}$
mass may be due to lattice spacing effects.

The SCALAR collaboration~\cite{Kunihiro:2003yj}, 
used unquenched lattice QCD,
with Wilson fermions and the Wilson gauge action,
to study the $0^{++}$ mesons.
At a single lattice spacing a $\sim$ 0.2 fm,
with $\overline{q}q$ interpolating operators
only, they obtain $m_{\overline{q}q} \sim $ $m_{\rho}$.
The lattice spacing dependence of this result needs
to quantified.

\begin{figure}
\centering
\includegraphics[%
  scale=0.4,
  angle=270,
  origin=c]{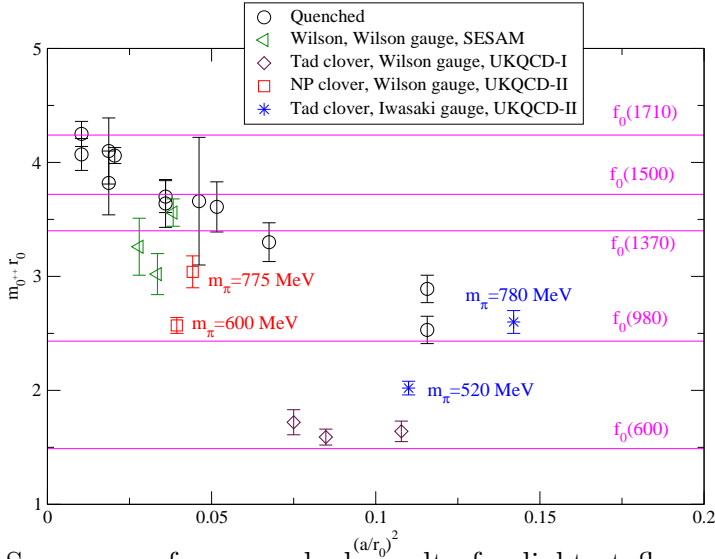}
\vspace{-2.5cm}
\caption{Summary of unquenched results for lightest flavour
singlet $0^{++}$ mesons from~\cite{Hart:2006ps}.
The unquenched results are from SESAM~\cite{Bali:2000vr},
UKQCD-I~\cite{McNeile:2000xx}, and UKQCD-II~\cite{Hart:2006ps}.}
\label{fig:unquenchGLUEBALL}
\end{figure}

In unquenched QCD, both glue and $\overline{q}q$ states
will couple to singlet $0^{++}$ mesons, so it is better
to do a variational fit with both types of operators as 
basis interpolating operators. 
The variational technique analysis of the 
singlet $0^{++}$ mesons was done by
Hart et al.~\cite{Hart:2006ps}.
A combined fit
to $0^{++}$ glue and $\overline{q}q$ interpolating
operators with two types of spatial smearing sources  was 
done. The calculation used the non-perturbative improved
clover action at a single lattice spacing~\cite{Allton:2001sk}.
Configurations from CP-PACS~\cite{AliKhan:2001tx}
with the Iwasaki gauge
action and tadpole improved clover action were also used 
in the analysis, because this calculation should be
less affected by lattice artifacts.
A summary plot of the results, in units of $r_0$ ($1/r_0 \sim$ 400 MeV) 
is in figure~\ref{fig:unquenchGLUEBALL}
(updated from~\cite{Hart:2006ps}).
%%%
%%%%%%%%%%%%%%%%%%%%%%%%%%%%%%%%%%
The data with the bursts and squares (with the pion masses
written near them)
in figure~\ref{fig:unquenchGLUEBALL}  
shows an additional reduction of the mass of the $0^{++}$ state over 
the pure glueball operators, as used by Hart and 
Teper~\cite{Hart:2001fp}.

Mathur at al.~\cite{Mathur:2006bs} recently
claimed to get a result for the mass of the
$f_0(600)$ ($\sigma$)
from quenched lattice QCD with pion masses
as low as 180 MeV. Using the
interpolating operator
$\overline{\psi} \gamma_5 \psi \overline{\psi} \gamma_5 \psi $
they obtain \mbox{$m_{f_0(600)} \sim 550$ MeV}. 
The key part of this work is a 
three state fit ($\pi(p=0)\pi (p=0)$ , $f_0(600)$,
$\pi(p=\frac{2\pi}{L})\pi (p=\frac{-2\pi}{L})$ using the Bayes adaptive
curve fitting algorithm~\cite{Chen:2004gp}. They studied the finite
volume effects to distinguish the signal for the resonance
from the $\pi\pi$ scattering states~\cite{Mathur:2004jr}.
Mathur et al.'s~\cite{Mathur:2004jr} calculation is discussed in 
slightly more detail in~\cite{McNeile:2007fu}. The effect of 
sea quarks on this calculation needs to be quantified.

There has also been a recent quenched QCD study~\cite{Suganuma:2007uv}
of light $0^{++}$ states with $qq \overline{q}\overline{q}$
interpolating operators that did not see resonant states in the quark
mass regime they explored.

In~\cite{Hart:2006ps} an attempt was made to compute the 
decay width for $f_0$ decay to two pions. Unfortunately 
much higher statistics will be required to obtain an accurate
value for that width.

\section{Conclusions}

There is still no consensus as to whether 
 $\overline{q}q$  operators in lattice 
QCD calculations are coupling to the 
$a_0(980)$ meson.
To clear up the many questions about the spectrum of the 
$0^{++}$ scalar mesons, unquenched lattice QCD calculations with
tetraquark interpolating operators are required.
There is ``some'' evidence that the flavour singlet $0^{++}$ 
interpolating operators,
in unquenched lattice QCD calculations, are coupling to states around
or below 1 GeV~\cite{Hart:2006ps}. Although a continuum
extrapolation is required for definite results.
Recent lattice QCD calculations that include the dynamics of the
sea quarks are working with light enough quarks
that the two meson decays of some scalar mesons 
are allowed~\cite{McNeile:2007fu,Michael:2007vn,Urbach:2007rt}.

\section*{Acknowledgments}

I thank Chris Michael for 
reading the paper. I thank the ETM collaboration for allowing
me to use preliminary results for the $a_0$ masses.

%%\bibliographystyle{h-elsevier2}
%%\bibliography{latt05}

%%%%%%%%%%%%%%%   Author and Subject Index
%%\printindex{author}{Author Index}
%%\blankpage
%%
%%\printindex{subject}{Subject Index}
%%\blankpage

\end{document}